\documentclass[12pt]{article}

\usepackage{amsmath}
\usepackage{amssymb}

\usepackage{graphicx}
\usepackage{subfig}
\usepackage{url}

\usepackage{cite}
\usepackage{color}

\usepackage{setspace}
\makeatletter
\renewcommand{\@biblabel}[1]{\quad#1.}
\makeatother

\date{\today}

\begin{document}

\begin{flushleft}
{\Large
\textbf{The hypothesis of urban scaling: 
formalization, implications and challenges}
}
\\
\vskip 0.5 cm
Lu\'is M. A. Bettencourt$^{1,\ast}$,
Jos\'e Lobo$^{2}$,
Hyejin Youn$^{1}$
\\
\vskip 0.5 cm
\bf{1} Santa Fe Institute, 1399 Hyde Park Rd, Santa Fe NM 87501, USA.
\\
\bf{2} School of Sustainability, Arizona State University, PO Box 875502
Tempe, AZ 85287, USA.
\\
\vskip 0.5 cm
$\ast$ E-mail: bettencourt@santafe.edu
\end{flushleft}

\section*{Abstract}
There is strong expectation that cities, across time, culture and level of development, share much in common in terms of their form and function. 
Recently, attempts to formalize mathematically these expectations have led to the {\it hypothesis of urban scaling}, namely that certain properties of all cities change, on average, with their size in predictable scale-invariant ways. The emergence of these scaling relations depends on a few general properties of cities as social networks, co-located in space and time, that conceivably apply to a wide range of human settlements. Here, we discuss the present evidence for the hypothesis of urban scaling, some of the methodological issues dealing with proxy measurements and units of analysis and place these findings in the context of other theories of cities and urban systems. We show that a large body of evidence about the scaling properties of cities indicates, in analogy to other complex systems, that they cannot be treated as extensive systems and discuss the consequences of these results for an emerging statistical theory of cities.


\pagebreak

\section*{Introduction}

There is a general recognition, common to many disciplines,  that cities regardless of their size, geography, time or culture share many underlying organizational, social and economic characteristics, and play similar functional roles in different human societies~\cite{Mumford,Hall,Bairoch,Smith_1}. A citizen of New York City will quickly understand how Tokyo works. She will also find a small town anywhere straightforward to navigate, if a little uneventful. When Cort\'es's men arrived in Tenochtitlan in 1519 (today's Mexico City) Bernal Diaz del Castillo famously described the city as spectacular for its scale ($\sim$200,000 people, one of the largest cities of the time) and wealth~\cite{Bernal}. But perhaps the true surprise should have been - given its independent development from old-world cites - how familiar it all was, in terms of its roads and canals, its public buildings and neighborhood organization and its markets and social life~\cite{Bernal}. The same could have been said of countless accounts of  travelers, historians and anthropologists.   There is a sense in which human settlements of ancient Mesopotamia and of modern developed nations share enough in common that the term "cities" can be used to meaningfully refer to entities separated by thousands of years of cultural, social and technological development~\cite{Mumford,Hall,Jacobs_Economy}.  All of this suggests that the functional role of cities in human societies, as well as some of the general aspects of their internal organization, may be universal: they may be expected to develop in urban systems that arose and evolved independently and hold across time, culture and level of technology. Cities, from this perspective, are variations - or perhaps better, elaborations - on a theme~\cite{PNAS,PLoS_One}.

The endeavor to discover general mathematical regularities of urban life, ''laws of cities" if you like,  is relatively new but increasingly possible given the growing availability of more and better data, and the multi-disciplinary scientific interest in the subject~\cite{Mumford,Jacobs_Economy,PNAS,Bettencourt_West,Wirth,Batty}.  Interest is also guided by the socioeconomic imperative of understanding cities in a fast urbanizing world~\cite{UN_report}.   The idea that cities - in some specific  sense - are  self-similar is what we call the {\it hypothesis of urban scaling}. In its strongest form it states that essential properties of cities in terms of their infrastructure and socio-economics are functions of their population size in a way that is scale invariant and that these scale transformations are common to all urban systems and over time. This means that there is no break in scale -- no minimum or maximum population size -- across which a city becomes no-longer a city:  say a village or a megalopolis. Cities of different sizes are not, however, the same because many important scaling transformations are non-linear~\cite{PNAS,PLoS_One,Carneiro_QQ}.  As a consequence, there are predictable and quantitative per capita savings in material infrastructure and increases in socio-economic productivity and innovation as a function of population size. Any urban system is ultimately rooted in material resources derived from food, energy and other basic materials but it is the connection of many (smaller) settlements with larger cities that drives the system as a whole to greater resource and economic efficiency and productivity, and permits increasing returns to the population scale of large cities in terms of innovation and wealth creation~\cite{PNAS,Jacobs_Economy,Fujita_Krugman_Venables}. These are ultimately the reasons why cities exist and can continue to grow. 

The urban scaling hypothesis, and the research that supports it, has non-trivial and subtle implications -- empirical, methodological and statistical -- while also overlapping with established research perspectives on cities~\cite{Fujita_Krugman_Venables,Christaller,Losch,Urban_Economics}. In the present discussion we address some of these implications and points of contact seeking to clarify and expand arguments presented elsewhere. Specifically, we consider the following questions:
\begin{enumerate}
  \item Is the urban scaling hypothesis synonymous with power-law behavior for urban observables?
  \item What is the relevant spatial unit of analysis?
  \item Over which range of population sizes do important attributes of cities exhibit scale invariance?
  \item What scaling proprieties can be expected to result from the mixture of local and national-level effects?
  \item How does the scaling hypothesis help us understand what type of complex system a city is?
  \item What are some of the methodological difficulties that arise when studying the properties of cities from current available data?
  \item How does the urban scaling hypothesis relate to and complement the existing body of work on urbanization and urban dynamics?
\end{enumerate}

Although the scaling hypothesis is a good  general description of many properties of cities across time and space,  there are, of course, plausible counter-arguments. It is possible, for instance, that below a certain population size a town may no longer have the functional properties of a larger city. Archeologists, for example, struggle with this distinction: When can we say that a settlement is urban? Does it require crossing a sharp threshold of density and population size, as it is often stated by census bureaus~\cite{census} and well known definitions of urbanism~\cite{Wirth}, or is there a more diffuse but non-linear continuum. Can cities exist without a developed urban system? Here, we discuss these concepts in greater detail, present new empirical evidence over a greater number of urban systems and a larger range of city sizes and propose several tests of the urban scaling hypothesis.  In so doing, we hope to sharpen the contribution made by research on urban scaling towards understanding the essential features of urban life.

\section*{Formalizing the hypothesis of urban scaling}

The statement that functional urban quantities should be scale invariant follows from the general observation that they are characteristic of population agglomerations of {\it all} sizes, from the smallest towns up to the largest mega-cities. By functional we mean the size of a place's economy, its amount of conflict, the extent of its infrastructure, its rate of innovation, etc. We do not mean, however, specific proxies for these quantities such as  amount of precious metals, murders by gunshot, number of patents, or of R\&D researchers,  etc. which are quantities specific to urban systems at a given time, particular technological context and level of socio-economic development. 

The requirement of urban scaling is that any average functional quantity, $Y$, is scale invariant, meaning specifically that
\begin{equation}
Y(\lambda N)/Y(N) = f(\lambda),
\end{equation}
where the function $f(\lambda)$ is independent of population size, $N$, but does depend on the arbitrary relative population size, $\lambda>0$. This simple assumption has been restated many times, for several decades and across several different disciplines~\cite{Nordbeck,Sveikauskas}, but it has not always been recognized that it implies the scaling relation
\begin{equation}
Y(N)=Y_0 ~ N^\beta,
\label{scaling_law}
\end{equation}
as can be verified by direct substitution. The constants in $N$, $Y_0$ and $\beta$, determine the scale of $Y$, more precisely $Y_0=Y(N=1)$, and the relative increase in the rate of $Y$ in terms of the rate of $N$, that is $\beta = d_t \ln (Y)/  d_t \ln (N)$. In general $Y_0$ is time dependent and varies from one urban system to another. The exponent $\beta$ is in general time independent (or slowly varying, at least) and takes similar values for similar quantities in different urban systems~\cite{PNAS,PLoS_One,Bettencourt_2012}, whether geographically or over time.
To prove that Eq.~(1) implies the scaling law in Eq.~(2), consider that $Y(N)$ is independent of the arbitrary parameter $\lambda$, which means:
\begin{eqnarray}
\frac{d Y(N)}{d\lambda}=0 \rightarrow \frac{1}{f(\lambda)} \frac{dY(\lambda N)}{d \lambda} - \frac {1}{f^2(\lambda)}\frac{d f(\lambda) }{d \lambda}  Y(\lambda N)=0.
\end{eqnarray}
Note that
\begin{equation}
\frac{ d Y(\lambda N)}{d \lambda} =  \frac{ d x}{d \lambda} \frac{ d N}{d x} \frac{ d Y(\lambda N)}{d N} = \frac{N}{\lambda} \frac{ d Y(\lambda N)}{d N},
\end{equation}
with $x=\lambda N$. Then, because $dN/N=d\ln N$, we can write
\begin{eqnarray}
\frac{d \ln Y(\lambda N)}{d \ln N} = \frac {d \ln f(\lambda)}{d \ln \lambda},
\end{eqnarray}
which we can be integrated to give
\begin{eqnarray}
Y( \lambda N) = \exp \left[ \int_0^{ \ln \lambda N}  d \ln N' ~  \frac{d \ln f(\lambda)}{d \ln \lambda} \right] = Y_0 e^{\beta \ln \lambda N}= Y_0 ~ (\lambda N)^\beta,
\end{eqnarray}
where $\beta \equiv \frac{d \ln f(\lambda)}{d \ln \lambda}$. The derivative in Eq.~(5) is independent of the specific value of $\lambda$, which implies that $f(\lambda)=\lambda^\beta$ Because of this property we can choose any scale $N=\lambda N'$ above and write the scaling law in its simplest form of Eq.~(2), as usual. This derivation reveals the mathematical assumptions necessary and sufficient to obtain scaling laws for cities. 

It is important to state clearly the meaning of Eq.~(1). It says that cities are self-similar in terms of scale $N$, so that regardless of their actual population size their average properties can be inferred from knowledge of those at another city, whose population is related to it by  a scale transformation $\lambda$. Over the range of population scale where it holds (and here it is assumed that it holds from $N=1$ to infinity, but see discussion below) this says that any functional property of a large city, organizational or dynamical, is already present in a small town and vice-versa, and that its quantitative prediction can be made via a non-linear scale transformation. 

So far we only discussed urban quantities, $Y$, as if they were deterministic.  It should also be clear that urban scaling is an {\it average} property of cities, so that it does not determine urban properties exactly but rather up to some level of uncertainty.  An ultimate theory of cities should provide predictions about urban indicator {\it statistics}, including the expected value of deviations from the mean scaling prediction and the correlations in time and space of these deviations.  The beginnings of a  {\it statistical} formulation of scaling is presented in detail in Ref.~\cite{Andres}, where it is shown how it emerges from the estimation of the conditional probability, $P(Y|N)$, as the expectation value of urban quantities $Y$, given a city's population $N$.  These issues are discussed in detail below.    Moreover, a theoretical framework that predicts the observed values of scaling exponents across many urban quantities from a few general properties of social and infrastructural urban networks has recently been proposed in Ref.~\cite{Bettencourt_2012}. This translates the expectation of scale invariance for urban indicators in terms of a few well known and well accepted generative principles and provides a detailed derivation of Eq. (2) and its parameters (such as $\beta=1+\delta$, $\delta \sim 1/6$, for socioeconomic rates). This theory also derives under what conditions the agglomeration advantages of cities may disappear, $\beta \rightarrow 1$.   

For the strong properties dictated by self-similar scaling to hold in practice for all urban systems we must ask that cities are defined in a way that is consistent across scales. The approach to these issues from the point of view of the scaling hypothesis sheds some light on difficult problems in the study of cities, such as the choice of unit of analysis, the extent of urban self-similarity, the nature of proxy quantities and the presence of truly local urban effects vs. non-local (national) effects, to which we now turn. 

\section*{What is the unit of analysis?}
A major practical difficulty in studying the properties of cities is the choice of unit of analysis.  Most official statistics pertain to somewhat arbitrarily defined administrative units, which, at some level, are not cities at all. Examples include counties or census tracts in the USA, several forms of local authority in the UK, prefectures in Japan, prefecture and district level cities in China, and municipalities in many European and South American nations. 

However, the appropriate definition of cities is functional, as strongly interacting, co-located social networks.  At the socioeconomic level there has been an increased effort to define cities in these terms. So, for example, we have the current definitions of micropolitan and metropolitan areas in the USA, which are in effect unified labor markets~\cite{census}, and similarly defined spatial units in Japan~\cite{Japan_MAs}, several Latin American nations~\cite{Andres}, the European Union (i.e., "Larger Urban Zones" \cite{Eurostat}), and so on.   The definition of these socioeconomic units requires measures of social interaction, or their proxies, which are difficult to obtain and analyze unambiguously. The U.S. Census Bureau utilizes a combination of population size, density and commuting flows data when defining metropolitan areas, but no simple algorithm is provided~\cite{census}. Nevertheless, the smallest integer unit that makes up any metropolitan area is a county. Similar ideas have been pursued in Europe  and Japan but the standardization of these definitions (especially in Europe) is still at a developmental stage.  Here, we want to emphasize and illustrate quantitatively some of the potential biases in urban indicators that can arise when units of analysis that are partial to a larger city or aggregates of cities are taken, instead of their appropriate definitions. We focus on the estimation of scaling exponents $\beta$. 

Consider a set of socioeconomic units with scale (e.g. population) $\{x_i\}$ and urban indicator $\{y_i\}$, with $i \in \{1,\ldots,n \}$. Then consider also the log-transformed variables $X_i=\ln x_i$, $Y_i=\ln y_i$.  Then, the scaling exponent $\beta$ can be computed in practice via, for example, the Ordinary-Least-Squares (OLS) estimator,
\begin{eqnarray}
\beta = \frac{\overline{XY} - \overline{X} ~ \overline{Y}}{\overline{X^2}- \overline{X}^2}.
\label{OLS-estimator}
\end{eqnarray}
Here, the bars over the symbols denote sample averages, so that, for example
\begin{eqnarray} 
\overline{XY} = \frac{1}{n} \sum_{i=1}^n X_i Y_i.
\end{eqnarray}
It is trivial to verify that, if we insert the relation $Y_i = \beta X_i$, we recover the value of the exponent from Eq.~(\ref{OLS-estimator}).

Consider now a situation where the units $x_i$, $y_i$ are correct urban definitions, but where we will further disaggregate them in terms of parts of a functional city according to the transformation where we take a datum $\{x,y\}$ and express it in terms of a set of points $\{w_i,z_i\}$ with $i=1,\ldots,m$. Under this operation of disaggregation, the log-transformed variables $W,Z$ become putative urban units in their own right (new cities) and the estimator for $\beta$ changes, by an amount $\delta \beta$ as
\begin{eqnarray}
&&\beta'=\beta+\delta \beta = \frac{\overline{XY} - \overline{X}~ \overline{Y}   + \delta_N}{\overline{X^2} - \overline{X}^2 + \delta_D} \\
&& \delta_N = \delta_{XY} - \overline{X} \delta_Y - \overline{Y} \delta_X - \delta_X \delta_Y, \quad \delta_D=\delta_{X^2} - 2 \overline{X} \delta_X - (\delta_X)^2,
\end{eqnarray}
where
\begin{eqnarray}
&& \delta_X = \frac{1}{n-1+m} \left[ m ( \overline{W} - \overline{X} ) + ( \overline{X} - X ) \right], \  \delta_Y = \frac{1}{n-1+m} \left[ m ( \overline{Z} - \overline{Y} ) + ( \overline{Y} - Y ) \right], \\
&& \delta_{XY} = \frac{1}{n-1+m} \left[ m ( \overline{WZ} - \overline{XY} ) + ( \overline{XY} - XY ) \right], \  \delta_{X^2} = \frac{1}{n-1+m} \left[ m ( \overline{W^2} - \overline{X^2} ) + ( \overline{X^2} - X^2 ) \right]. \nonumber
\end{eqnarray}
Here the sample averages over the new variables $W$ and $Z$ are taken over their range, that is, for example
\begin{eqnarray}
\overline{W} = \frac{1}{m} \sum_{i=1}^m W_i.
\end{eqnarray}

This expression, though exact, is not very transparent.  We can see what the change in the exponent $\beta$ is, typically, by expanding Eq.~(9) to first order in small $\delta_N$ and $\delta_D$, to obtain
\begin{eqnarray}
\delta \beta \simeq \frac{ \delta_N - \beta \delta_D}{\overline{X^2} - \overline{X}^2}.
\end{eqnarray}
This expression can be further simplified, if, without loss of generality, we choose to have started with variables $X_i$ and $Y_i$ such that $\overline X=\overline Y=0$. Then, collecting leading terms, we obtain
\begin{eqnarray}
\delta \beta \simeq \frac{m (\beta_{WZ} -\beta) \overline{W^2} + (\beta-\beta_{XY}) X^2}{(n-1+m)\overline{X^2}},
\end{eqnarray}
where
\begin{eqnarray}
\beta = \frac{\overline{XY}}{\overline{X^2}}, \quad \beta_{XY} = \frac{XY}{X^2}, \quad \beta_{WZ} = \frac{\overline{WZ} - \overline{W} ~ \overline{Z} }{\overline{W^2} - \overline{W}^2}.
\end{eqnarray} 
Thus, we see immediately that if all variables scale with the same exponent, $\beta$, the correction vanishes, $\delta \beta=0$.  A small correction may occur if the variables $X,Y$, to be disaggregated, are atypical in terms of scaling, that is if $\beta_{XY}$ is very different from $\beta$.  The most import change, however, results from the leading correction for large $m$, if the variables $W_i$, $Z_i$, do not scale, or do so with a very different exponent from $\beta$.   In general, this correction can have either sign. If $W_i$ and $Z_i$ are anti-correlated, meaning, for example,  that parts of the city with greater population have lower incomes than expected from the scaling relation and vice versa, this will reduce the expectation for $\beta$. This could result, for example, if income was reported by place of work vs. place of residence in cities organized in terms of central business districts and residential suburbs. Or by strong income inequality in terms of low density rich neighborhoods compared to high density poor parts of the city. Thus, strong heterogeneity inside the city as well as specific forms of reporting can bias exponent estimates, and naturally result in washing out agglomeration effects. Only when the strongly mixing components of the city are aggregated together can one see the city for what it is in terms of its socioeconomic performance. 

Likewise, it can be shown that when several cities are aggregated together the exponent $\beta$ will tend to become more linear as would be expected if a putative larger interacting population fails to realize its full agglomeration effects in terms of socioeconomic output or savings in material infrastructure. To show this, consider the aggregation of two cities into one, so that  $x_+=x_1+x_2$ and $y_+=y_1+y_2$.  Then define, as above, $X_+ = \ln x_+$ and $Y_+=\ln y_+$. We find that, under this aggregation, 
\begin{eqnarray}
&& \delta_{XY} = \frac{1}{n} \left( X_+ Y_+ - X_1Y_1 -X_2 Y_2 \right), \quad \delta_{X^2} = \frac{1}{n} \left( X_+^2 - X_1^2 -X_2^2 \right), \\ 
&& \delta_{X} = \frac{1}{n} \left(  X_+  - X_1 -X_2\right), \quad \delta_{Y} = \frac{1}{n} \left(  Y_+ - Y_1 -Y_2 \right),
\end{eqnarray}
so that, to first order in $1/n$,
\begin{eqnarray}
&& \delta \beta \simeq \frac{1}{n \overline{X^2}} \left[ (\beta_+ - \beta) X_+^2 + (\beta-\beta_{12}) (X_1^2 +X_2^2) \right],
\label{final_deltabeta_agg}
\end{eqnarray}
where $X_+^2 \beta_+ = X_+ Y_+$ and $(X_1^2 +X_2^2) \beta_{12} = X_1Y_1 +X_2 Y_2$.   
This last expression can be simplified further when we assume that the original variables scale exactly, $\beta_{12}=\beta$, so that we are left with the first term in Eq.~(\ref{final_deltabeta_agg}), which can be written as 
\begin{eqnarray}
\delta \beta = \frac{X_+}{n \overline{X^2}} \left[ \ln \beta +(1-\beta) \ln (x_1 +x_2) \right].
\end{eqnarray}
It is now easy to see that, for sufficiently large cities, the magnitude of the second term is always larger, so that the correction to $\beta$ becomes negative if $\beta>1$ and positive if $\beta<1$, thus shifting the true exponent towards unity, and underestimating agglomeration effects.  In particular if we write $\beta=1+\delta$, with $\vert \delta \vert <<1$ \cite{Bettencourt_2012}, this leads to 
\begin{eqnarray}
\langle \delta \beta \rangle \simeq - \delta \frac{X_+}{\beta_X} \left[ \ln (x_1 +x_2) -1 \right],
\end{eqnarray} 
which takes the {\it opposite} sign of the deviation, $\delta$, in the exponent from unity as long as  the sum of populations $x_1+x_2 > e$. 

Thus, we have shown that the choice of unit of analysis is crucial in estimating the correct urban scaling exponents.  In general, if cities are under or over aggregated this can lead to the underestimation of the magnitude of urban agglomeration effects.  

\section*{The extent of self-similarity}

There is by now ample evidence that important properties of cities increase, on average, faster (socioeconomic superlinearity) or slower (material infrastructure, sublinearity) than city population size~\cite{PNAS,Batty,PLoS_One,Bettencourt_2012,Shalizi}. These properties hold across time and different urban systems, even those at very different levels of development, though with different baselines, $Y_0$. The issue is what is the best quantitative characterization of these size dependences and in particular over which range of population do they manifest scale invariance.

Most presently available datasets of urban indicators have a relatively small range of scales, at least when compared to systems in physics or biology. Even in the largest urban systems cities are limited to tens of million people ($10^7$) and data are usually not available for units below $N\sim10^3-10^4$, leaving us with only about 3-4 orders of magnitude. Urban indicators span a similar range. Over this range it is often the case that our putative scale invariant functions, Eq.~(\ref{scaling_law}),  can be modeled in terms of other {\it scale-dependent} functions, such as $Y \sim N \log N$,  or other functions with more parameters~\cite{Shalizi}.  The fact that these other functions of $N$ can fit urban data well (but not better) than power-laws in most observed cases, see Figs 1-3, has been used as an argument against the scale invariance of urban indicators~\cite{Shalizi}. However, this merely means that other models may be viable explanations of the scaling behavior of urban indicators, along with scale invariant functions. In science, a model can never be proven correct, of course, but models can be excluded if they make not only empirically wrong predictions but also imply counterfactual consequences or if they contradict fundamental theoretical expectations. Here we discuss some of these issues. 

It is always the case that any power law, especially with a power close to 1, can be approximated in terms of a sum of powers in logarithms, as is obvious from the Taylor expansion
\begin{equation}
Y(N)=Y_0 N^\beta= Y_0 N e^{(\beta-1) \ln N} = Y_0 N \left[ 1+ (\beta-1) \ln N + \frac{1}{2}((\beta-1) \ln N)^2 + O \left[ ((\beta-1) \ln N)^3 \right] \right].
\label{log-expansion}
\end{equation}
As long as $ (\beta-1) \ln N <<1$ the first few terms in the expansion give an excellent approximation to the original function. Because $\ln N \sim 15$ and $\beta-1 \sim 1/6$ we should expect $(\beta-1) \ln N$ not much larger than unity, see Figure 1A, where the expansion in Eq. (\ref{log-expansion}) is shown to produce an excellent match to data.  Alternatively, {\it by introducing a scale in the problem}, $N_{\rm m}$, the function
\begin{eqnarray}
Y(N)=C N \ln \frac{N}{N_{\rm m}}
\label{log}
\end{eqnarray}
typically improves the empirical fit at the cost of having implicitly assumed that as $N$ becomes small $Y$ vanishes, i.e. $N\rightarrow N_{\rm m}$,  $Y(N) \rightarrow 0$.  For cities smaller than $N_m$ the logarithmic fit predicts {\it negative} and fast diverging urban indicators.  

The observation that these functions can often fit the data as well as the power law was made recently for GDP and personal income in US metropolitan areas~\cite{Shalizi} and is indeed true empirically in most cases, see Figs 1-2.  As we already discussed above,  our reason for not considering Eq.~(\ref{log}) as a viable fit to urban data in our early analyses~\cite{PNAS,PLoS_One} was due to its strange consequences as it implies that common urban indicators, such as personal income, wages, GDP and all others would vanish in towns below a critical size (see {Figs 1-2}) and presumably be zero or need to be specified in some additional manner below $N_m$. Also, empirical fits with this form predict that $N_m$ would be quantity specific, vary over time and from one urban system to another, see Figs. 1-3.   Figure 1A shows the scaling of personal income with population for US cities larger than 10,000 people (micro- and metropolitan statistical areas) in 2006. This shows that both the power law function and the scale-adjusted logarithm, Eq.~(\ref{log}), fit the data equally well  and that the scale $N_m \sim 1$.  Personal income, as we shall show below is a problematic quantity as it mixes truly urban dynamics with nation-wide economic transfers. As such, its scaling exponent $\beta=1.06$ is unusually low, which allows these two functions to coincide over a larger population range.  Fig. 1B shows patents filed in the same cities. While it remains true that the scale-adjusted logarithmic function and the power law fit the data equally well, the former now predicts a sharp nose dive in patent production for small cities that would vanish at $N_m \sim 200$ people. Similar effects occur for other urban indicators in other urban systems: Fig 2A shows the number of murders in Brazilian Metropolitan areas and smaller municipalities. We fitted both the power law function and the scale-adjusted logarithm to metropolitan areas and investigated how these predictions generalize to smaller municipalities. While the power law performs reasonably well and its exponent for metropolitan areas is consistent with that for the aggregated data set, the scale-adjusted log function generalizes poorly, fails to predict the rate of violence in small municipalities, which it would have vanish at the critical scale of $N_m=14,454$ people.  Figure 3B shows the scaling of income in Japanese Metropolitan areas: again, while both types of functions work well among large cities, the adjusted scale logarithm would predict vanishing income for $N_m\sim257$ people.    

Because we find these statements counter-factual, and because the scale-adjusted logarithmic function does not derive from theory to the best of our knowledge or fit the data manifestly better, we believe the hypothesis of urban scaling in terms of a scale invariant function, predicted by theory \cite{Bettencourt_2012}, remains at this point the better explanation.   Nevertheless, the explicit observation of small settlements provides a testable prediction that can distinguish these models. In this vein, some of us have recently applied the hypothesis of urban scaling to archeological evidence from the Prehispanic Basin of Mexico~\cite{BOM}.  In this study of one of the most complete urban systems of an early major civilization, we find human settlements as small as 10 people and about 1 hectare in land area. Urban systems across four major cultural periods spanning about two millennia show power-law scaling of settled area compatible with modern cities and general theoretical predictions~\cite{Bettencourt_2012}.  Although more evidence for the properties of small settlements is desirable, we take this as another indication that the hypothesis of urban scaling applies very generally. 

Another general issue deals with the potential dynamical consequences of superlinear power-law scaling as a driver of urban growth.  This can lead to a finite-time singularity, which is sometimes approximately observed~\cite{Kremer_1993,PNAS}.   A logarithmic function potentially shows a qualitatively different behavior~\cite{Shalizi}.  To test these ideas we present below the analytical solution that follows for the analogue of the urban growth equation  proposed in \cite{PNAS}.  Consider that some quantity $Y$, which scales faster than linearly is seen as driving the growth dynamics of a city, while we assume here for analytical tractability that costs scale linearly (but see supplementary materials in \cite{PNAS} for a complete analysis). Then the growth equation is
\begin{equation}
\frac{dN}{dt} = Y(N) - B N.
\label{growth-equation}
\end{equation}
The solutions for a power-law function, $N_\beta(t)$, was given in \cite{PNAS}; 
\begin{eqnarray}
N_\beta(t) = \left[ \frac{Y_0}{B} + \left(N_0^{1-\beta} - \frac{Y_0}{B}  \right) e^{-B(1-\beta)t}  \right]^{1/(1-\beta)}. 
\label{power_law_solution}
\end{eqnarray}
The solution for the logarithmic function, $N_{\ln}(t)$, is
\begin{eqnarray}
N_{\rm ln} (t) = N_{\rm m} \exp \left[ \frac{B}{C} + \left(  \ln \frac {N_0}{N_{\rm m}} -   \frac{B}{C} \right)~e^{C t} \right],
\label{log_solution}
\end{eqnarray}
both obeying the initial condition $N(t=0)=N_0$. Thus the logarithmic function is qualitatively different from the power law in that it does not create a pure finite-time singularity: Instead this solution grows in time like an exponential of an exponential. However, these two solutions are very hard to distinguish unless one is very close to the (unphysical) singularity, which is not generally possible. 

The true singularity, where $N \rightarrow \infty$, produced by the superlinear driving term, starting at $N(t=0)=N_0$,  occurs at a finite time $t_c (N_0)$
\begin{eqnarray}
t_c^\beta=\frac{1}{(\beta-1)B} \ln \frac{1}{1 - \frac{B}{Y_0} N_0^{1-\beta}}\simeq  \frac{1}{(\beta-1) Y_0 N_0^{\beta-1} }.
\label{singularity}
\end{eqnarray}
Short of this singularity, both driving terms predict an acceleration of the population growth rate $r=\frac{1}{N}\frac{d N}{dt}$ in time and as a function of the initial population size, $N_0$.  If we compute the characteristic time that it takes to reach a certain maximum growth rate $r^*$, from  Eq.~(\ref{power_law_solution}) and (\ref{log_solution}), we obtain in each case
\begin{eqnarray}
t_\beta^* = \frac{1}{(\beta -1 )B} \ln \frac{1-\frac{B}{r^*+B}}{1 - \frac{B}{Y_0} N_0^{1- \beta} } , \qquad t_{\ln}^* = \frac{1}{C} \ln \frac{r^*}{C \ln \frac{N_0}{N_m} - B }.
\end{eqnarray}
The first expression reduces to Eq.~(\ref{singularity}) in the limit of large rate $r^*$ and large $N_0$, which shows how $t^*$ becomes shorter with larger $N_0$ ~\cite{PNAS}. This property remains true even when these limits are not taken, though its expression is more complicated. Likewise, expression for $t^*_{\ln}$ under log-driven growth, even in the absence of a true finite time singularity, also shows the same properties. We write it as 
\begin{eqnarray}
 t_{\ln}^* = \frac{1}{C} \ln \frac{ r^*}{y(N_0) -B},
 \end{eqnarray}
where $y(N_0) = Y(N_0)/N_0$, which shows that as $N_0$ becomes larger the initial condition in the growth equation approaches  a fixed $r^*$ in a finite and decreasing time (Note that the initial condition is such that $y(N_0)>B$, for all $N_0$). Thus,  the general observable properties of the growth rate remain essentially the same, regardless of the use of a scale invariant driving function, or of an adjusted scale logarithm that mimics it. It would be interesting to investigate in the future if there are growth curves for cities from which the expression for the driving term can be estimated in a sufficiently robust way that hypotheses about different driving terms can be tested. In reality, however, the growth of cities in time is more complex than that given by Eq.~(\ref{growth-equation}), because of a more complex cost structure~\cite{Bettencourt_2012} and because the time variation of the pre-factors, such as $Y_0$, cannot be ignored. 

\section*{Mixed quantities: urban vs. national effects}
A practical issue with estimating scale invariant relations is that available data does not always refer to quantities that are truly local to each city. One example is personal income. Personal income is a combination of several components~\cite{personal_income}, such as wages, profits from investments and transfers. Transfers refer to government payments that effectively re-distribute wealth from richer (larger) cities to poorer (smaller) ones. These nation-wide transfers have several effects on the statistics of this urban indicator. First, the fit is unusually good, with very small outliers. Second, its scaling exponent is only weakly superlinear, $\beta=1.06-1.07$. 
These effects are partly due to the fact that personal income as given is not a pure measure of the cities economic performance but partially the result of urban system wide redistributions. Figure~3 shows the scaling of net earnings\footnote{As defined by the US Bureau of Economic Analysis, "Net earnings is earnings by place of work (the sum of wage and salary disbursements, supplements to wages and salaries, and proprietorsÕ income) less contributions for government social insurance, plus an adjustment to convert earnings by place of work to a place-of-residence basis"} and of transfers. We see that net earnings are more superlinear $\beta=1.12$ but transfers (received) are slightly sub-linear with an exponent $\beta=0.96$.

What scaling properties, if any, can we expect from the mixture of these components? In general the answer is that the sum of two power-law functions is not a power law. However, in practice, we may expect that the sum may behave {\it approximately} as a power law, provided that the two exponents are sufficiently similar. To see how this can be, consider the function
\begin{equation}
Y(N) = A N^a + B N^b.
\label{mock_function}
\end{equation}
As we have already noted above, the exponent of any function $Y(N)$ can be estimated via the quantity
\begin{eqnarray}
\beta=\frac{d \ln Y}{d \ln N}= \frac{1}{Y} \frac{d Y}{d \ln N} = \frac{A a e^{a \ln N} +B b e^{b \ln N} }{ A e^{a \ln N} +B e^{b \ln N}}.
\end{eqnarray}
Then, we obtain, for small $\epsilon \ln N$, $\epsilon =a-b$,
\begin{equation}
\beta= \frac{ A a + B b}{A + B} + \frac{A B}{(A+B)^2}  \epsilon^2 \ln N +O(\epsilon^3 \ln^2 N).
\label{mixed_exponent_estimator}
\end{equation}
This formula generalizes trivially to more components in the sum. Thus,  provided the exponents are similar the (approximate) exponent for the sum is well predicted by the weighted average of the exponents of each additive scaling function (first term in Eq.~(\ref{mixed_exponent_estimator})), with the first correction being controlled by the small quantity
$\epsilon^2 \ln N$.  Consequently to a very good approximation the resulting exponent is the weighted average of the two underlying exponents. For example, if $A=B$ then the resulting exponent will be the average $\beta=(a + b)/2$, to leading order. 

Another related issue is the shape of the distribution of deviations away from scaling. 
These are the residuals of a log-log fit, which in Ref.~\cite{PLoS_One} we defined as Scale Adjusted Metropolitan Indicators (SAMIs),
\begin{equation}
\xi_i(t) = \log \frac{Y_i}{Y_0(t) N_i(t)^\beta}.
\label{SAMI}
\end{equation}
In Refs.~\cite{PLoS_One,Andres,Markus} we briefly addressed the shape of the statistical distribution of $\xi$ and showed that it was reasonably well described in terms of a Gaussian (that is a log-normal in the original variables), at least at the tails. Here, we simply want to emphasize that a mixture of variables hence distributed will not have a similar distribution, and may show skewness. 

Consider then a mixture of $Y_1$ and $Y_2$ quantities, each of them obeying on average a scaling law with exponents $\beta_1$ and $\beta_2$. We will assume that these two variables are  log-normally distributed around the scaling relation, for simplicity. While the sum of the log variables, which have Gaussian distributions, is also Gaussian 
\begin{eqnarray}
\log (Y_1) + \log(Y_2) \sim {\cal N}(\mu,\sigma)
\end{eqnarray}
it is not true that the $\log$ of the sum of log-normals is Gaussian.  Several approximations have been derived in the literature to extract the properties of the distribution of sums of correlated or independent lognormal variables~\cite{sum_lognormals,sum_approx_tail,skewed_lognormals}. Most of these match other distributions to the measured parameters of the sum distribution~\cite{sum_approx_tail,skewed_lognormals}. It has been shown that the tails of the sum distribution are often approximately log-normal~~\cite{sum_approx_tail,skewed_lognormals}, but the body of the distribution deviates from a simple log-normal shape. In particular it often shows skewness, so that a {\it skewed} log-normal distribution provides a good general fit to the sum~\cite{skewed_lognormals}.  These properties of residuals' statistics, which may occasionally appear, may be taken as signs that the underlying quantity may be an additive mixture of more fundamental urban indicators and that it may combine urban effects with nation-wide dynamics, such as transfers, in the case of personal income.

\section*{Cities are non-extensive complex systems}

Here, we discuss a conceptual and methodological issue underlying the study of cities across different disciplines. Specifically, should we think of cities as extensive systems, in the sense used in statistical physics, with constant city size-independent densities (per capita quantities) including population density, wages, crime?   Or should we, instead, approach the study of cities from their global properties across the city (total wages, total GDP, total crime) and the implicit expectation that densities within the city are non-intensive, highly variable and must therefore be approached statistically?

To be fair the general answer to this question is well known: all complex systems are non-extensive and their (obvious) densities are non-intensive; that is they depend {\it non-linearly} on measures of total system size.   This means, for example, that when we divide a complex system in parts we cannot generally expect the average properties of each part to be similar to each other. We also find that the properties of the parts, including its basic elements (such as people and locations in a city), depend on the size of the entire system.  Think of an organism or a brain or an ecosystem, where different parts perform different functions. Cities too, show extreme spatial and individual heterogeneity: there are rich and poor neighborhoods,  there are business districts, which are almost exclusively dedicated to jobs and trade and there are suburbs, which are mostly residential. There are parts of the city dedicated to certain types of business, manufacturing, etc.   Likewise, there are poor and rich individuals, people with radically different economic and social expertise, motivations, ethnicity, etc. The city only makes sense when all these parts are put together as an interacting social system~\cite{Jacobs_Economy,Bettencourt_2012}: There is no such thing as a representative average place or average person inside the city. There is, however, a general sense of the global socioeconomic and spatiotemporal fabric of cities. 

The question of unit of analysis is fundamental because it deals with targets of explanation and theory development.  For example, in urban economics one attempts to explain (average) per capita quantities, such as wages per worker  or crime rates. However, it is uncommon in economics to develop {\it statistical} microscopic theories.  As a counterpoint to this approach, when studying complex systems we often start with the bulk properties of a system because of the expectation that the whole is `more than the sum of its parts', that is that the properties of individuals are a reflection of system-wide dynamics~\cite{more_is_different}. These global properties are what is directly measurable and are usually simpler than individual attributes from which the statistics of densities - not just their means - are derived.  In our previous and current work we have advocated the latter approach~\cite{PNAS,PLoS_One,Andres}. Here we provide some of our rationale in greater detail.

The difference between extensive and non-extensive systems and of the type of theory that is appropriate to describe them is best illustrated via a few examples.  In simple physical systems intensive quantities are well defined -- meaning that they are truly system size independent -- when a system's total properties are proportional to the system's size.  In this case densities are ratios of extensive quantities, which are linear function of system size,  as their trivial size dependence cancels out leaving us with constants that characterize the system. The simplest familiar example is a(n ideal) gas in a container whose bulk properties are proportional to the volume and to the number of particles therein. The average energy per particle (i.e. the temperature) and the force per unit area (pressure) are then simple functions of each other and of the volume and number of particles. Extensive systems have no shape: They are well characterized by homogeneous densities and stretch to fill the available volume, provided exogenously. Thus, for these simpler systems, extensive or intensive variables are trivially interchangeable (because $\beta=1$) and the latter provide well defined average properties of its microscopic elements. The statistics of individual elements is provided in turn from distributions that maximize entropy, subject to constraints arising from the average value of intensive quantities (like particle density and kinetic energy), e.g. the Maxwell Boltzmann distribution of particle velocities in a gas at temperature $T$.  

Now, contrast these properties with those of cities: Cities are localized in space and have definite shapes. Though we measure directly total quantities, such as total GDP, total crime or population, we know that their properties - imagined as continuum densities over space - vary dramatically from one location within the city to another and among different people, as is emphasized e.g. by crime hotspots~\cite{heterogeneity} or other population density maps at night or during the day.   From this perspective recent work in urban economics and urban scaling has established the non-extensivity of cities as infrastructure and socioeconomic properties do not vary proportionally (linearly, $\beta=1$) to surface area or population. Furthermore, unlike gas molecules individual members of an urban community display tremendous variability with respect to creativity, entrepreneurship, productivity,  propensity for violent behavior, etc~\cite{Markus}.  

How then should we start to characterize a city?   Non-extensive systems in physics are the result of long-range forces, such as electromagnetism or gravity: canonical (equilibrium) thermodynamics generally breaks down as a global theory in these cases, or must at least be handled under specific restrictions~\cite{Lynden-Bell}.   Not incidentally, attractive forces are often invoked to explain several phenomena relating to cities.  For example,  gravity models have been quite successful at accounting for population movements, such as migration flows~\cite{gravity_models}. Moreover, agglomeration effects invoked in urban economics and economic geography~\cite{Fujita_Krugman_Venables} are a form of attractive interaction between people or firms, even if the quantitative form of this force is not explicitly known. Likewise, the movements of people inside cities are dictated by social interactions that afford them an income, and other opportunities. Thus, we should think of cities as social systems that result from attractive forces. 

What can be said about cities from the perspective of the statistical properties of systems assembled by attractive interactions?   One familiar result of gravitational forces is the creation of stars, which are particle reactors held together by the balance of radiation pressure leaving the star and gravity compressing it. Cities in urban economics and regional science are also thought to be the result of the balance between centrifugal (congestion, high land rents, crime) and centripetal (beneficial social interactions, agglomeration effects) forces~\cite{Fujita_Krugman_Venables}.   The resulting structures, in stars and in cities,  are not globally rigorously stable but can persist for long times in a state of dynamical {\it local} equilibrium. This means, in general, that each location or person within the system is the result of the dynamical balance of attractive and repulsive forces, whose magnitude is a function of their position relative to all other elements in the system.  This often creates mono-centric geometries, but not necessarily.  In this state, locally attractive and repulsive forces equilibrate each other leading to a spatial profile of varying density, pressure, and temperature. There is no sense of the typical density or temperature in a star as it varies dramatically from the center (at nuclear densities) to the periphery (vacuum of outer space). Stars, like cities, must then be understood in terms of global quantities and the conditions and constraints that they impose within their structure.  By promoting faster interaction rates, however, both larger stars and cities create total outputs that scale superlinearly with their size: for example the total power emitted by starts (luminosity) is also a superlinear function of their total mass, albeit with different exponents from cities due to the nature of radiation and its transport~\cite{Bettencourt_2012}. 

Statistical theories of even simple systems governed by attractive forces remain, to some extent, a general open problem~\cite{PADMANABHAN}.  However, statistical mechanics of these systems can be developed based on intensive quantities that are either place and time dependent~\cite{Prigogine} and/or judiciously constructed by removing average size effects.  When this is done, we may expect that the resulting statistics become simpler and may belong to the exponential family of distributions. 

\section*{Statistical theories of cities}
A statistical theory of cities is necessary to eventually account for individual and collective variability in and across urban areas.   Statistical theories, in the sense of theoretical physics, are uncommon in geography, and especially among the social sciences, including economics (with the notable exception of finance). Here, we discuss some initial attempts at characterizing the statistical properties of urban quantities and their connection to scaling relations. 

As we pointed out in the preceding section, the main challenge to formulating a statistical theory of urban quantities is dealing with the size dependence of most densities, expressed usually as per capita quantities. In Ref.\cite{PLoS_One}, we showed that certain simple scale invariant quantities can be constructed using scaling. This procedure was further explored in Ref.~\cite{Andres} where we gave a more precise statistical meaning to urban scaling relations and established their connection to Zipf's law for the size distribution of cities. Here we place these findings in a wider context and discuss some of what remains to be done on the path towards a statistical theory of cities.

First, the statistics of urban indicators can be approached directly, in terms of the statistics of $Y$, or indirectly via associated quantities that are scale independent.  It is observed empirically that the distribution of $Y$, for example for murders~\cite{Andres}, is very broad, and is generally well described by a power law (Pareto) distribution.  On the other hand,  the conditional distribution of $P(Y|N)$ is more localized.  Formally, this means that population size has information on the values that $Y$ is likely to take. The estimation of these distributions is often made difficult by the type of proxy data available. For example, small enough cities may show, on a given year, zero, one, two murders, patents, etc.  Does this mean that scaling breaks down in these cases?  The answer is no, not necessarily at least, but that the expectations for the number of $Y$ must be treated statistically, that is,  there may be a small probability that a small city will have one event per year, but this {\it probability}, not the detailed outcome, remains non-zero and a scaling function of $N$.

We showed recently that this implies that the correct statistical interpretation of scaling laws is as  expected values of $Y$, {\it given} $N$.   Moreover, this conditional statistics is naturally in the lognormal family~\cite{Andres}.  To see this consider that 
\begin{eqnarray}
P(Y) = \sum_N P(Y|N) P(N),
\end{eqnarray}
where $P(N)$ is the probability of a city of population size $N$, related to Zipf's rank-size distribution. Indeed, with $P(N)$ given approximately by a power law, $P(Y)$ will also be Pareto distributed if $P(Y|N)$ is lognormal~\cite{Andres}. 

In practice for quantities that are granular and can be zero at each realization, we estimate instead $P(N|Y)$, and use Bayes' relation to compute $P(Y|N)$, 
\begin{eqnarray}
P(Y|N) = \frac{P(N|Y) P(Y)}{P(N)},
\end{eqnarray}
which is distributed also as a lognormal, if $P(N)$ and $P(Y)$ are Pareto and $P(N|Y)$ is lognormal, as it is for murders in several Latin American urban systems~\cite{Andres}.  The type of distribution observed for other variables remains an open problem, but we should note already that the two conditional distributions may take more complicated forms if the city size distribution deviates from Pareto, or vice-versa. 

An alternative approach, inspired by the general theoretical considerations discussed in the previous section, is to create  {\it ab initio} quantities that are invariants of city size and, as such, provide suitable densities for which we may expect well behaved statistics to emerge.  Some of these invariants are suggested by theory~\cite{Bettencourt_2012}: for example the product of any social output per capita times the volume of infrastructure per capita is, on average, city size independent.  But more generally we can build intensive quantities by explicitly removing their size dependence~\cite{PLoS_One}, such as for the SAMIs $\xi$, Eq.~(\ref{SAMI}).  

The correspondence between the statistics of $Y$ and those of $\xi$ is just what one may expect in terms of transforming the original lognormal statistics into more familiar statistical mechanics. Besides subtracting the conditional mean of $Y$ given $N$, which is the scaling law, the logarithmic transformation results in Gaussian statistics for $\xi$.  Thus, this approach maps the somewhat exotic statistics of non-extensive urban indicators, $Y$, to canonical statistics of well defined densities, $\xi$. The $\xi$ appear to always have bounded variance, so that Gaussian statistics may be expected, at least in the absence of additional constraints. Exploring these statistics more extensively remains an important problem for future research. 

From the perspective of the approach sketched here one can obtain a general {\it statistical} characterization of many urban indicators. For example, we have recently shown that a Cobb-Douglas production function is the expected value for the economic output of a city, conditional on its population size, and determined its statistical behavior~\cite{Lobo_Bettencourt}.   The eventual goal of a theory of cities should be to determine the form and value of urban indicator statistics, and specifically of the variances of size-independent densities, which are quantity specific, as can be seen in Figs.~1 and 2.  Such a theory should also explain the spatial and temporal correlations between densities and why these statistics, when expressed in their original variables, appear to be the result of multiplicative random processes~\cite{Montroll}.

\section*{Population size, agglomeration and urban hierarchies}
Although methodologically novel, the hypothesis of urban scaling has many points of contact with existing literature on cities in the fields of archaeology, anthropology, sociology and economics, as well as with the extensive historiographical treatments of cities and urbanization. This is not the place to attempt a comprehensive literature survey, but a few remarks on the empirical and conceptual overlaps between the urban scaling hypothesis and other research traditions serves to strengthen the empirical underpinnings of the hypothesis, locate it within the myriad of efforts to understand cities and point the way to research extensions.

The role of population size in spurring organizational and technological change has been an important formative idea in economics since at least Alfred Marshall's account of industrial districts~\cite{Marshall_1890,Arrow_1962,Simon_1977,Simon_1986,Boserup_1981,Lee_1987,Lee_1988,Kremer_1993,Lucas_1993}. Population size (in cities) has been proposed many times as the main general transformative phenomenon in ancient and modern human societies, see, for example, \cite{Naroll_1956,Carneiro_QQ,Carneiro_2000,Ember_1963, Dumond_1965,Diamond_1978,Diamond_1997,Shennan_2002,Klasen_Nestmann_2006,Algaze_2008,Kline_Boyd_2010,Boyd_Richerson_Henrich_2011}.   In all these cases it is not just population size that matters, but the resulting social dynamics of interaction in a larger pool of people that requires and promotes new organizational and technological solutions.  Memorably, Jane Jacobs in {\it The Economy of Cities}~\cite{Jacobs_Economy} defined a city as a population agglomeration that through its organization and innovation is able to generate (endogenously) its own economic growth. 

This foundational work led to many efforts to quantify the economic properties of cities as a function of their population size. The most important conceptual framework of modern urban economics is based on the idea of {\it agglomeration economies}, which are a form of externality that results in a set of benefits that firms and individuals obtain when locating near each other\footnote[1]{As Lucas (\cite{Lucas_1988}, p.39) asks: "What can people be paying Manhattan or downtown Chicago rents for, if not for being near other people?"}~\cite{Sveikauskas,Rosenthal_Strange_2001,Puga,Glaeser_2008,Glaeser_Resseger_2010}. A dense interacting co-location of people and firms is, of course, exactly what a city is~\cite{Bettencourt_2012}.  A large empirical literature in urban economics has attempted to measure the economic effects of urban agglomeration, when controlling for other factors, such as the level of education in the population, age, sector composition, etc. Thus, this empirical approach measures {\it net} agglomeration effects, which are almost always characterized in terms of power-law relations vs. population. Over the last forty years, exponents characterizing these net scaling relations have been obtained for several measures of economic output, in different nations and controlling for different factors. Scaling exponents have been estimated in the range 3-8\% ~\cite{Rosenthal_Strange_2004}. 
These results imply that the measurements of the dependence of gross socioeconomic output on population size, that is of the dependence of $Y$ on $N$ in the absence of control variables, must be larger, as observed by scaling analysis. One issue with measuring net agglomeration elasticities (exponents), using many of the common control variables, is that these controls are naturally themselves population size dependent, following similar power-law scaling relations~\cite{PNAS}.  It is also difficult to consistently include the same controls in different nations and over time and factors like human capital remain difficult to measure objectively, which has been cause for some controversy as to the magnitude of its importance as an input factor. 

Besides the measurement of net or gross agglomeration effects,   urbanists, regional scientists and economic geographers have attempted to characterize the general mechanisms that lead to productivity increases through social interactions.  In somewhat different ways, authors as diverse as Marshall~\cite{Marshall_1890}, Mumford~\cite{Mumford}, Jacobs~\cite{Jacobs_Economy,Jacobs_1984}, Pred~\cite{Pred_1977}, Braudel~\cite{Braudel_1979}, Bairoch~\cite{Bairoch}, Hall~\cite{Hall}, Florida~\cite{Florida_2004}, Glaeser~\cite{Glaeser_2011} and Kennedy~\cite{Kennedy_2011} all emphasize in their writings the advantages conferred by larger urban population size.  Specifically, a  larger population allows for a more extensive division of labor and specialization, engenders greater diversity, fosters the generation, exchange and recombination of ideas, makes it possible for scale economies to manifest themselves  and facilitate the building of physical and social infrastructure. Duranton and Puga~\cite{Duranton_Puga,Puga}, have attempted to systematize these effects in terms of three categories of mechanisms, namely {\it sharing}, {\it matching} and {\it learning}. Sharing refers primarily to the joint use of indivisible assets, such as a hospital or a stadium. Matching refers to a more productive allocation of workers to jobs, for example, while learning deals with processes of information transfer between individuals and firms.  These categories of effects provide a useful organizational framework bridging several economic processes but they may not capture all socioeconomic phenomena leading to increases in productivity with size.  Another example from sociology refers to Claude Fisher's subculture theory~\cite{subculture} (see also Wirth's influential early essay~\cite{Wirth}), which argues that larger populations promote new (unconventional) behaviors, from deviance to innovation, reflected in a greater number of new subcultures, created by small groups of people with particular interests in common. Note that this effect is not necessarily economic, and deals in general with the creation of new information and culture through recognizable new social collectives.   This rich literature provides many examples of the socioeconomic effects that can be stimulated by city size, but, contrary to scaling~\cite{PNAS,Bettencourt_2012}, it does not formalize these detailed phenomena quantitatively, in terms of the structure and dynamics of social networks that are predictive of observed agglomeration effect exponents.   

Population size figures prominently in another long-standing research tradition on cities, namely urban hierarchies. Cities do not exist in isolation as they are linked to each other through the exchange of goods, people, capital and information.   The starting point for urban hierarchy models -- and central postulate of central place theory~\cite{Christaller} --  is the observation that larger cities have a more diverse set of industries and occupations than smaller places. Hierarchy means specifically that the set of activities and occupations in larger places include those found at smaller settlements, but not the reverse. Many models of urban hierarchies are based on `central place theory', under which cities function as `central places' providing greater value-added goods to the surrounding areas~\cite{Christaller}.  Variations include L\"osch's analysis of hierarchical centers based upon market areas for industrial firms~\cite{Losch} and the Tinbergen-Bos model, which distributes industries among cities of different sizes according to their relative economies of scale \cite{Tinbergen_1961, Bos_1965}.   Evidence for the existence of functional hierarchical patterns in urban systems has been well documented~\cite{Mori,Bettencourt_Samaniego_Youn} and they have been shown to hold not only for urban systems within nations but also across national boundaries~\cite{Angel_2012}.  The link between population size, urban industry and occupational structure in terms of a hierarchy of greater value added services implies that wage or productivity decompositions, in which the average wage or GDP per worker of an urban area is `explained' as a result of its industry composition (see, for example, \cite{Drennan_2002,Shalizi}) is a redundant exercise. Only in the absence of a size hierarchy of functions and productivity, that is, in a putative but non-existent urban system where each city would fully specialize in different industries, would such a decomposition be potentially explanatory.  The ideas of central place theory have continued to be elaborated and, in conjunction with concepts of urban agglomeration, introduced through assumed forms for urban production functions with increasing returns~\cite{Fujita_Krugman_Venables}, constitute the core of modern economic geography.  From this perspective, urban hierarchies can indeed be derived from a more microscopic theory of social interactions in cities that is the goal of urban scaling~\cite{Lobo_Bettencourt,Bettencourt_2012,Bettencourt_Samaniego_Youn}.  

The main mystery that remains regarding the structure of urban systems deals, in our opinion, with their dynamics over time.  Present theories of regional economics (in analogy with central place theory) provide a means of deriving spatial equilibria (distributions of people and economic activity) across cities in an urban system, but they do not tell us how they evolve in time. Likewise, urban scaling research has been primarily concerned with the value of urban exponents, which are approximately constant over time, but not with the time dependence of the pre-factors, $Y_0$, that describe how the urban system evolves as a whole.   
%
The constancy of exponents implies that larger cities do not, on average, grow faster than cities in smaller size groups and the average growth rates of cities of all population sizes tend to be the same~\cite{Gabaix,Angel_2012,Angel_2010}. This independence of growth rates, not only of population but also for other socioeconomic properties, is the proximate cause of Zipfian size distribution for city sizes across the urban system~\cite{Gabaix}. However, how exactly this equilibration happens, such that cities large and small hold the same appeal and grow economically at the same pace despite their productivity differentials remains an important test to a science of cities as well as to a general theory of socioeconomic development~\cite{Jones_Romer}.   

\section*{Discussion}

In this manuscript we strived to clarify the hypothesis of urban scaling, discuss some of its potential limitations and methodological issues  and highlight how scaling phenomena reflect the properties of cities as general complex systems. We have also highlighted what are, in our opinion, the main outstanding problems on the way to a science of cities as a unified field, recognizable across disciplines, from physics to the full spectrum of the social sciences. 

First, in its strongest form, the hypothesis of urban scaling states that functional properties of cities, such as their level of conflict, economic productivity and material infrastructure should all vary in a scale invariant way from the largest cities to the smallest towns within an urban system. The expectation that this picture emerges as empirically true is grounded in recent theory~\cite{Bettencourt_2012}, but should be further tested by measurements on all scales, as small settlements remain typically unaccounted for in most datasets. Nevertheless, we take it as factual, but yet to be quantified -  that even the smallest settlements, including those studied by anthropologists~\cite{Kline_Boyd_2010}, archeologists~\cite{BOM} and sociologists,  have elements that functionally find correspondences in larger modern cities. These generic arguments then establish the only possible scale invariant function - a power law - as the preferred description of cities across scales. Any other function, even if it fits available data equally well, introduces a population scale thresholds below which this property vanishes. Such scales, when introduced in fitting functions, are observed to vary empirically from quantity to quantity, across urban systems and over time, leading to what, in our opinion, are potentially counterfactual results.  This case, however, must ultimately be investigated and settled empirically and is in our opinion an interesting direction for future work.

Second, several methodological issues remain unsettled in the empirical study of cities related to spatial aggregation, statistical interpretation of the data and limitations of proxy measurements. At present, there are no full-proof algorithms for defining the socio-spatial limits of a city or even a fully generalizable characterization of urban life,  though the concept of functional cities as interacting populations has firmly emerged as the correct idea. Nevertheless, this remains hard to delimit  unambiguously in practice, as efforts at the pan-european level have demonstrated over the last few years, for example. The choice of an arbitrary administrative unit (counties, municipalities, etc) often distorts the estimation of agglomeration effects and modifies the conclusions of scaling analysis.  We have also shown how certain urban metrics (e.g. personal income) may manifest deviations from typical expected scaling behavior as the result of mixing truly urban (local) dynamics with counteracting national effects. These issues can in principle be addressed and corrected for but they require understanding of the nature of proxy variables and their limitations,

Finally, we have shown that cities are not extensive systems and, as a consequence, that densities (per capita or per area urban indicators) are necessarily system size dependent and heterogeneous, typically varying widely within a city. Additionally, there is significant heterogeneity among the constituent units of cities, namely individuals, households and businesses.  These properties are the necessary consequence of the generic dynamics of agglomeration that characterizes the formation of cities. The fact that densities are not direct observables but constructs of extensive quantities should make us prefer city wide total measurements as the primary characterization of cities. To us, and in analogy  to successful quantitative approaches in other disciplines,  this situation suggests that intensive quantities should be approached statistically, as the result of city-wide constraints (total population, area, transportation technology, etc) and not the reverse. Of course, in many statistical analyses of urban metrics the average information contained in per capita measurements constructed over the entire city is equivalent to that in total quantities. We expect that these issues will be settled as statistical theories of cities develop further.


For their central role in contemporary human societies we expect that cities will continue to be the increasing focus of intense and multidisciplinary analyses.  We echo the observation of authors such as Paul Romer or Richard Florida that cities are the principal socio-economic organizational units of the 21st century, much like firms were in the 20th century.  Whatever form a general theory of cities will ultimately take, its contours in terms of the advantages of urbanization and its potential limits are beginning to take shape. Human societies generally develop organizational structures, economies, innovations  and ways to manage conflict that develop across scales in ways that are approximately self-similar as the result of general social network interaction dynamics.  We believe that the hypothesis of urban scaling expresses these self-consistent changes in a synthetic and testable way that may help reveal the general mechanisms underlying large scale human sociality.

\section*{Materials and Methods}

\section*{Acknowledgments}
We thank Clio Andris, Marcus Hamilton and Geoffrey West for discussions.  This research is supported by grants from the Rockefeller Foundation (no. 2011 SRC 108), the Bill \& Melinda Gates Foundation (no. OPP1076282), the James McDonnell Foundation (no. 220020195), the National Science Foundation (no. 103522), the John Templeton Foundation (no. 15705) and the Bryan J. and June B. Zwan Foundation.

\bibliography{template}

\pagebreak 

\begin{figure}[!ht] 
  \centering
 \mbox{
\subfloat[]{    \includegraphics[width=0.5\textwidth]{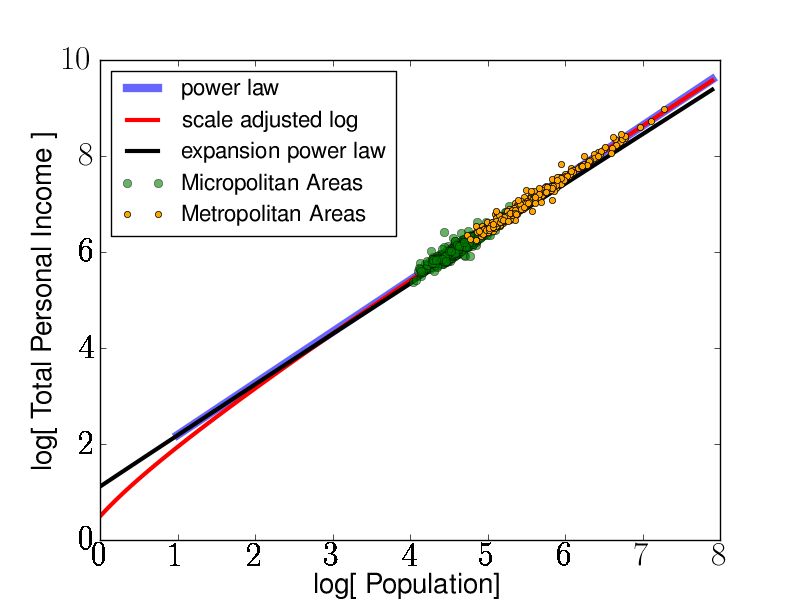} }
\subfloat[]{        \includegraphics[width=0.5\textwidth]{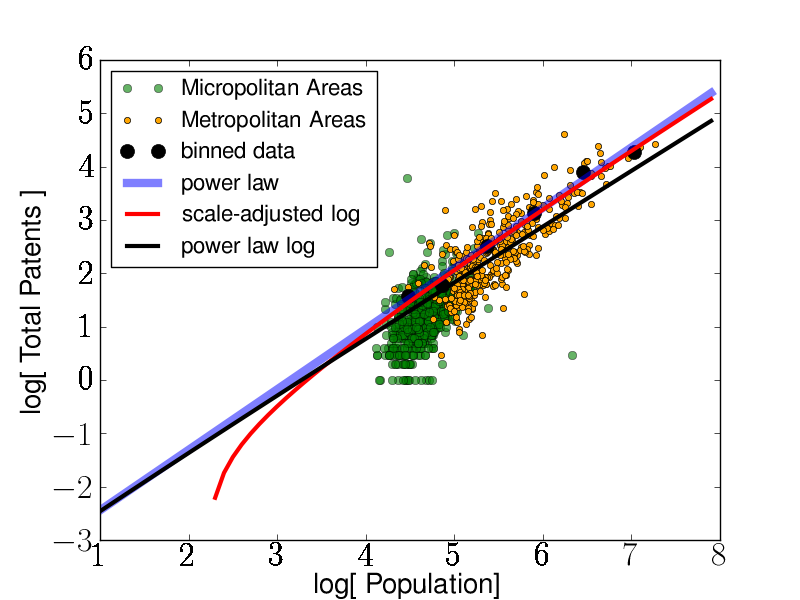}}
  }
\caption{Scaling of personal income and patents in US micro and metropolitan areas. (a) Total personal income in 2006. Metropolitan areas are shown in orange, with micropolitan areas shown in green.  Lines show the best fit to a scaling relation $Y=Y_0 N^\beta$ (blue), with $\beta=1.07$ (95\% CI [1.06,1.09], $R^2=0.66$), the series expansion to the power law in terms of logs to first order (black), and a fit to the scale-adjusted logarithm $CN \log N/N_m$ (red). The scale-adjusted logarithm and the power law fits are indistinguishable in terms of quality of fits where data are available. (b) Total number of patents filed in US metro and micropolitan areas in 2006. Black dots indicate logarithmically binned data for visual guidance. The power law fit has exponent $\beta=1.13$ (95\% CI [1.01,1.26], $R^2=0.66$). Again, the adjusted-scale logarithm and power law fits are indistinguishable where data are available but the former predicts that a city with less than 200 people would have zero patents and that smaller cities would display a negative number, a seemingly counterfactual result of the introduction of a scale $N_m=200$ in the problem.}
  \label{Figure 1}
\end{figure}

\begin{figure}[!ht] 
  \centering
 \mbox{
\subfloat[]{    \includegraphics[width=0.5\textwidth]{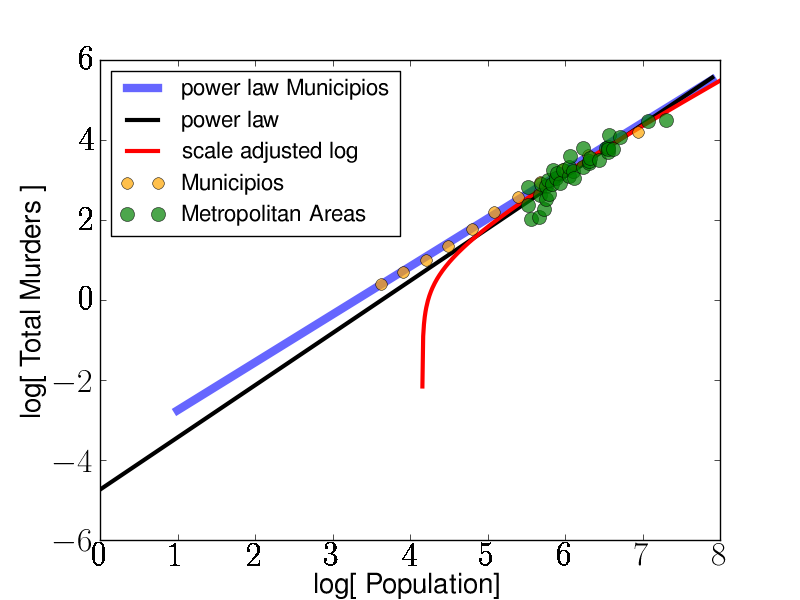} }
\subfloat[]{        \includegraphics[width=0.5\textwidth]{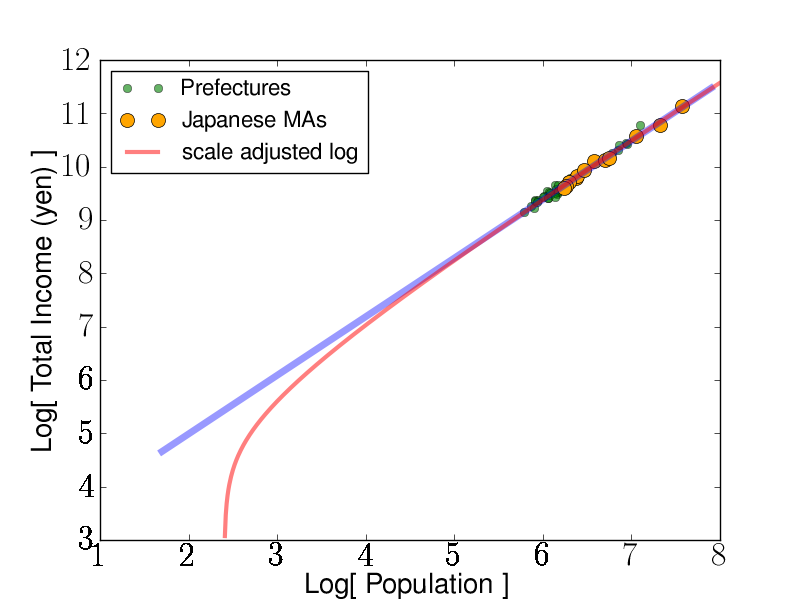}}
  }
\caption{Scaling of murders in Brazilian Cities and Income in Japanese Metropolitan Areas (a) Total number of murders per year in Brazilian Metropolitan Areas (green) and non-metropolitan Municipalities (orange), binned logarithmically (yellow).  Lines show the best fit to a scaling relation for metropolitan areas plus municipalities, $Y=Y_0 N^\beta$ (blue), with $\beta=1.20$ (95\% CI [1.15,1.25], $R^2=0.96$), a power law fit to metropolitan areas only (black) with $\beta=1.30$ (95\% CI [1.12,1.49], $R^2=0.85$), and a fit to $CN \log N/N_m$, or scale adjusted-scale logarithm (red), which dives to zero at the critical city size of 14,454, which is manifestly wrong given the data on municipalities. (b) Total income in Japanese metropolitan Areas (MAs) in 2006 (orange) and prefectures (green).  The power law fit (blue line) has exponent $\beta=1.10$ (95\% CI [1.04,1.16], $R^2=0.99$). The adjusted-scale logarithm and power law fits are indistinguishable where data are available but the former goes to zero at a critical city size of $N_m \sim 257$ people.}
  \label{Figure 2}
\end{figure}

\begin{figure}[!ht] 
  \centering
 \mbox{
\subfloat[]{    \includegraphics[width=0.5\textwidth]{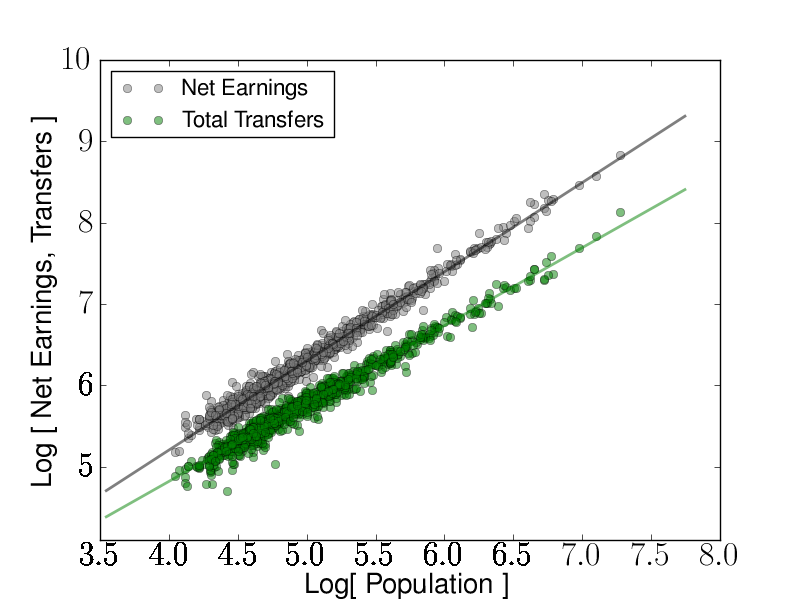} }
\subfloat[]{        \includegraphics[width=0.52\textwidth]{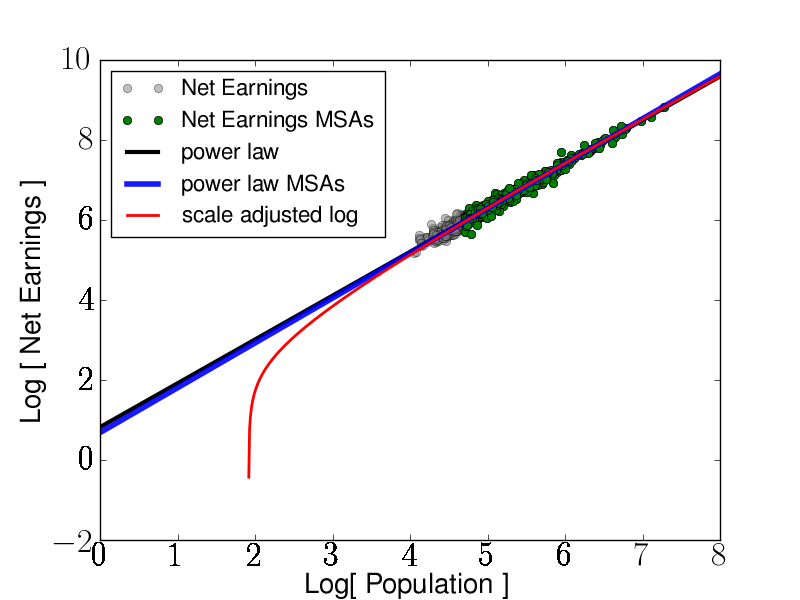}}
  }
\caption{Scaling of components of personal income in US cities(a) Total net earnings and transfers for micro and metropolitan areas in the US in 2007.  Note how these two quantities, the former generated locally in the city and the latter the result of national policy, scale differently. When added together they result in approximate scaling for personal income with a lower exponent than generally expected from urban scaling theory~\cite{Bettencourt_2012}. Lines show the best fit to scaling relations  with $\beta=1.10$ (95\% CI [1.05,1.15], $R^2=0.98$) for net earnings and $\beta=0.96$ (95\% CI [0.91,1.01], $R^2=0.97$) for transfers. (b) Net Earnings in US micro and metropolitan areas. The power law fit to the entire range (micros + metro areas) is very close to that for metros only (blue). The best fit exponent for metropolitan areas is $\beta=1.12$ (95\% CI [1.10,1.14], $R^2=0.98$). On the other hand the scale adjusted logarithm predicts a city of 83 people with zero net earnings, which seems counterfactual.}
  \label{Figure 3}
\end{figure}


\end{document}